\documentclass[12pt]{article}

\usepackage{amsmath} 
\usepackage{amsthm}
\usepackage{amssymb}
\usepackage{ulem}
\usepackage{xcolor}
\usepackage{hyperref}

\newcommand{\be}{\begin{equation}}
\newcommand{\ee}{\end{equation}}
\newcommand{\bea}{\begin{eqnarray}}
\newcommand{\eea}{\end{eqnarray}}

\newcommand{\EEA}{\end{eqnarray}}

\begin{document}

\thispagestyle{empty}

\title{Emergence of conformal properties in Finite Grand Unified Theories via reduction of couplings}
\date{}
\author{Myriam Mondragón\thanks{email: myriam@fisica.unam.mx}~ and 
  Luis Enrique Reyes Rodríguez\thanks{email: enriquerrdz@estudiantes.fisica.unam.mx }\\[0.3cm]
{\small
 Instituto de Física, 
Universidad Nacional Autónoma de México, }\\
{\small Ciudad de México, C.P. 04510
Mexico}
}

{\let\newpage\relax\maketitle}

\begin{abstract}

Zimmermann’s Reduction of Couplings (RoC) method is a powerful tool for addressing the problem of the excess of parameters in a field theory, as it yields relations among couplings that are invariant under the renormalization group. Its usefulness becomes particularly evident when constructing predictive supersymmetric GUT models that are free of UV-divergences to all orders. Within this scale-invariant framework, we show that a SUSY model satisfying the conditions of all-loop finiteness exhibits a conformal regime induced by superpotential operators compatible with the RoC. In the soft-breaking sector, the method was shown to lead to a set of scale-invariant relations between the soft couplings and the parameters of the dimensionless sector, among which a sum rule for the scalar masses is particularly notable. These relations closely resemble the typical AMSB relations, while avoiding the tachyonic mass spectrum thanks to the sum rule. Based on this observation, we provide new evidence for a connection between the reduction of parameters in the dimensionful SSB sector and the emergence of an anomaly-mediated (AMSB-like) pattern, under the assumption that the finite Grand Unified Theory is connected to an effective $N=1,~d=4$ Weyl-invariant SUGRA theory. In this process, we find that an all-order finite model with this property requires a specific form of the Kähler potential, whose structure coincides with that studied in no-scale supergravity scenarios.

\end{abstract}
\maketitle

\section{Introduction} 
\label{intro}

The Standard Model (SM), although very successful in describing the fundamental interactions, leaves many questions unanswered, expressed in its large number of free parameters.  The addition of symmetries, although a natural way to relate different parameters, often introduces even more through enlarged symmetry groups,  which lead to more particles and/or interactions, and different ways of breaking said symmetries.  A complementary and powerful approach to study relations among parameters was introduced by Zimmermann  \cite{Zimmermann:1984sx,Zimmermann:2001pq}, which consists on systematically finding renormalization group invariant (RGI) relations among  parameters,  relating them to a fundamental or primary one. This is known as the Reduction of Couplings (RoC) method, which combined with the addition of symmetries has proven very successful in reducing the number of free parameters and to construct highly predictive models.

The notion of a theory that remains well defined and complete at high energies—namely, a theory free from UV divergences—rests on a theoretical framework in which the appearance of infinities in physical observables is interpreted as signaling the presence of additional degrees of freedom or interactions at energy scales beyond the unification scale, and possibly approaching the Planck scale. Consequently, the consistency of a fundamental theory, i.e. UV complete, valid at such energies requires that it be finite. The search for such theories aligns directly with  Zimmernan's reduction of couplings method \cite{Zimmermann:1984sx}. This makes it possible to construct a high-energy theory that not only benefits from the reduction method but also remains free of ultraviolet divergences. All of this is based on the search for parameter relationships that are invariant under the renormalization group.
It is also the basis for looking for the absence of infinite divergences, or all-loop finiteness, in a theory, where the $\beta$ functions vanish to  all-loops, which is only possible in a supersymmetric setup \cite{Lucchesi:1987ef}.

The study of finite theories arises in the context of $N=1$ supersymmetric Grand Unified Theories (GUTs) under the idea of the absence of divergences at scales larger than the unification scale, where all interactions are expected to appear in a single scheme. The conditions for one-loop finiteness for  $N = 1$ SUSY field theories with $SU(N)$ gauge group were first studied in \cite{Rajpoot:1984zq,Rajpoot:1985aq}, and the   necessary and sufficient for finiteness at the two-loop level in refs.~\cite{Parkes:1984dh,Jones:1984cx}.  As already mentioned, all-loop finiteness can be achieved through the RoC method \cite{Lucchesi:1987ef}, guaranteeing that the theory is free of ultraviolet divergences.
The effect of this absence of divergences is not directly observable at low energies, where the unification gauge group is broken, but rather indirectly, since the finiteness requirement gives relationships between the gauge and Yukawa couplings at the unification scale \cite{Lucchesi:1987ef}, which in turn lead to specific values for the masses at low energies. 
Thus, theories are considered that include the Standard Model as a low-energy version of a more fundamental theory that exhibits greater symmetry at higher scales. Since supersymmetry (SUSY) is a requirement for a finite theory, the successful theories studied so far assumed that the $N=1$ SUSY finite GUT breaks down to the Minimal Supersymmetric Standard Model (MSSM), and then to the SM, which led to successful predictions first for the mass of the top quark  and then for the Higgs boson mass, prior to their respective discoveries \cite{Kapetanakis:1992vx,Heinemeyer:2007tz,Heinemeyer:2012sy}. For reviews on the Reduction of Couplings method and its applications to particle physics see refs.~\cite{Heinemeyer:2014vxa,Heinemeyer:2019vbc}.

For its part, supersymmetry breaking through the inclusion of the soft-breaking Lagrangian introduces,  in general, of the order of one hundred new independent dimensional parameters. In refs~\cite{Parkes:1984dh,Jones:1984cu,Jack:1994kd} the conditions for finiteness in the soft breaking sector were found through RGIs, and it was shown that also in this sector one-loop finiteness implies two-loop finiteness. On the other hand, by  implementing Zimmermann’s method it is possible to reduce the soft breaking SUSY  parameters by constructing relations among them and the parameters of the dimensionless sector of the supersymmetric theory, valid at the unification scale and above \cite{Kubo:1996js,Kobayashi:1997qx} in Gauge-Yukawa unified theories, as well as in finite theories. These RGI relations are highly relevant, since they ensure that in a finite theory even the soft breaking terms  are scale invariant  above the $M_{\text{GUT}}$ scale \cite{Jones:1984cu,Jack:1994kd}. Moreover, in ref.~\cite{Kobayashi:1997qx}  a two-loop  sum rule among the soft scalar masses in finite theories was derived, which is crucial to obtain viable phenomenology. 

Another consequence of this reduction in the supersymmetry-breaking sector is that the relations mentioned above are consistent with supersymmetry breaking induced by conformal anomaly effects (AMSB) \cite{Randall:1998uk}, which is somewhat surprising since no explicit assumptions about the type of mediation are made in their derivation. What we will show here is that Zimmermann’s reduction method acts as a bridge connecting Finite  Unified Theories (FUT) and conformal-invariant supergravity theories through an ``AMSB-like'' breaking mechanism.
It is worth mentioning that, although there were already indications of this connection between the RGI relations, which are a necessary condition for the Roc method, and the AMSB mechanism \cite{Jack:1999aj}, this is the first study to address its origin in detail in the context of all-loop finite Grand Unified Theories and supergravity.

There are examples of   $N=1, ~d=4$
string vacua that lead to effective finite field theories (namely at two-loops, see for instance~ \cite{Hamidi:1984ft,Ibanez:1998xn,Lawrence:1998ja}), as well as criteria to obtain finite theories from branes \cite{Hanany:1998ru} and more recently from quiver gauge theories \cite{He:2018gvd}, although a phenomenologically realistic model is yet to be found.  Also, there are several examples of predictive Grand Unified Theories finite at two-loops and even at all-loops \cite{Jones:1984qd,Kapetanakis:1992vx,Kubo:1994bj,Ma:2004mi,Heinemeyer:2007tz}, but a direct connection to a more fundamental theory was lacking. In this work, we will provide a detailed description of the conditions that make a supersymmetric unification theory finite at all orders in perturbation theory, using the RoC methodology, as well as the immediate consequences of imposing such conditions. Among the most interesting implications we found is the existence of a regime in  parameter space that exhibits superconformal invariance in the dimensionless sector, if the finiteness conditions are satisfied. Then,  assuming that the finite theory is  related to an $N=1$ conformal (Weyl-invariant) supergravity theory and using the RoC method, we show how the AMSB-like mechanism emerges. We found that in this case the Kähler potential has the same form as in no-scale supergravity theories, providing a natural link to inflationary scenarios \cite{Kallosh:2013yoa,Ellis:2015xna,Antoniadis:2024ypf}. 

The paper is organized as follows, after this brief introduction,
in Section II we  review the Reduction of Couplings Method, to then introduce the notion of finiteness based on the RoC method in Section III. In Section IV, the relation between the conditions of the theorem that characterizes finite supersymmetric theories and the conformal invariance associated with the dimensionless sector of these models is discussed in detail. Finally, Section V describes the reduction method when supersymmetry breaking is introduced, and provides a derivation/detailed explanation of why, upon reducing the dimensionful soft-breaking couplings, the effects of the conformal anomaly (AMSB-like) dominate over the other contributions associated with supersymmetry breaking, and we show the connection of these finite theories to conformally invariant SUGRA models. In the last section we present a discussion and our conclusions.

\section{Reduction of Couplings}
The main idea of the RoC method is to express all dimensionless parameters of a given theory in terms of a basic parameter, which is called the primary coupling. To achieve this, Renormalization Group Invariant (RGI) relations among couplings are assumed, which have the general form 
$$\Phi (g_1,\cdots,g_A) ~=~\mbox{const.}~, $$
and should satisfy the partial  differential equations:
\be
\mu\,\frac{d \Phi}{d \mu} = {\vec \nabla}\Phi\cdot {\vec \beta} ~=~
\sum_{a=1}^{A}
\,\beta_{a}\,\frac{\partial \Phi}{\partial g_{a}}~=~0~,
\ee
where $\beta_a$ are~the~$\beta$-functions of $g_a$.
The above ~equivalent~to the following ordinary differential equations,  called Reduction Equations~(REs)
\cite{Zimmermann:1984sx,Oehme:1984yy,Oehme:1985jy}:
\be
\beta_{g} \,\frac{d g_{a}}{d g} =\beta_{a}~,~a=1,\cdots,A-1~,
\label{redeq}
\ee
where $g$ and~$\beta_g$ are~the primary coupling and its corresponding $\beta$-function.
To express all couplings in terms of a single one $g$,  $A-1$ relations of the form $\Phi (g_1,\cdots,g_A)~=~\mbox{const.}$ are needed. However, the general solutions to Eqs.~(\ref{redeq}),  contain as many integration constants as the number of equations themselves, which means that  we just have  traded an integration constant for each ordinary renormalized coupling, and
these general solutions would not be reduced ones. The crucial requirement to solve this issue is  to demand that the above~REs admit power series solutions:
\be
g_{a} = \sum_{n}\rho_{a}^{(n)}\,g^{2n+1}~,
\label{powerser}
\ee
which are a set of special solutions and preserve~perturbative~renormalizability. Such an ansatz fixes the corresponding integration constant in each of the
REs and picks up a special solution out of the general one, thus leaving a truly reduced system.
The uniqueness~of these solutions can be decided at one-loop~level
\cite{Zimmermann:1984sx,Oehme:1984yy,Oehme:1985jy}.

In most cases the ``complete'' reduction of the parameter space described above proves to be unrealistic and a ``partial'' reduction (in which some parameters are left independent) is applied \cite{Kubo:1985up,Kubo:1988zu}.

\section{Finiteness on a Supersymmetric Grand Unification
Theory}

There are various proposals for mechanisms by which the gauge and Yukawa sectors within a unification theory can  be related. However, some of these proposals face consistency issues regarding the values of the third family fermion masses compared to experimental data \cite{Decker:1979cw,Chaichian:1995ef}, as well as the existence of additional infrared divergences \cite{Pendleton:1980as,infrared}. The implementation of Zimmermann's method of reduction of couplings (RoC)\cite{Zimmermann:1984sx} represents a natural way to relate different couplings to a primary one in perturbation theory. In particular, it is possible to achieve gauge-Yukawa unification in the context of a Grand Unified Theory (GUT). At a certain scale, the theory will be characterized by a single independent parameter. Under this assumption, the possibility of constructing  Finite Grand Unified models has been studied by analyzing the dimensionless parameters of a supersymmetric gauge theory, namely the Yukawa couplings and the gauge coupling. 
The conditions that guarantee that the dimensionless couplings of $N=1$ SUSY GUTs are  finite to one- and two-loops were derived already many years ago \cite{Parkes:1984dh,Jones:1984cx}, and are incompatible with models which contain a $U(1)$, like the MSSM. Thus, larger unification groups have to be considered.  

For our purposes, we will consider only the study of supersymmetric quantum field theories $N=1$. The treatment of theories with extended supersymmetry ($N=2,4$) can be found in references \cite{Grisaru:1986wj,Kazakov:2007dy}. Therefore, consider an $N=1$ supersymmetric field theory with a gauge symmetry given by a simple group $G$ \footnote{In general, this treatment can also be applied to semisimple groups formed by products of the form $SU(N)^k$ \cite{Ma:2004mi}.} and free of anomalies. Its superpotential is given by
 \begin{equation}
 \label{eq:sup3}
W=\frac{1}{2}m_{ij}\Phi^i\Phi^j+\frac{1}{6}C_{ijk}\Phi^i\Phi^j\Phi^k,     
 \end{equation}
 where $\phi_i$ transforms in the irreducible representation $R_i$ of $G$. The one-loop gauge and Yukawa beta functions of the renormalization group are constrained in their form thanks to the non-renormalization theorem for the superpotential \cite{Seiberg:1994bp}. This theorem states that the superpotential does not receive perturbative corrections. Moreover, due to the holomorphy in the fields $\phi_i$, severe restrictions are imposed on the beta functions for the dimensionless couplings $C_{ijk}$. Thererefore, the evolution equation under the renormalization group of a generic parameter will be proportional to the parameter itself and to the anomalous dimension $\gamma^i_j$ of the matter fields. With this considerations, the beta functions take the form 
 \begin{equation}
 \label{eq:bet}
 \beta^{(1)}_g=\frac{g^3}{16\pi^2}\Big[\sum_iT(R_i)-3C_2(G)\Big],
\end{equation}

\begin{equation}
\label{eq:BetYuk}
     \beta_{ijk}^{(1)}=C_{ijl}\gamma^l_k+C_{ikl}\gamma^l_+C_{jkl}\gamma^l_i,
\end{equation} 
and in the same way, the one-loop anomalous dimension of the superfields are given by

\begin{equation}
\label{eq:gamma}
    \gamma^{(1)i}_j=\frac{1}{32\pi^2}\Big[C^{ilm}C_{jlm}-2g^2C_2(R_i)\delta^i_j\Big].
\end{equation}

It is easy to see that if equations (\ref{eq:bet}) and (\ref{eq:gamma}) are made zero, then all one-loop beta functions of the theory must necessarily be zero as well. According to refs.~\cite{Parkes:1984dh,Jones:1984cx}, this would correspond to necessary and sufficient conditions to guarantee finiteness in a theory at least to first and second order. The direct consequence of this requirement is a constraint under the gauge group parameters and representations, as well as a direct relationship between all the dimensionless parameters of the theory, and it is similar to that expected using the method of RoC
\begin{align}
\label{eq:C100}
    \sum_iT(R_i)&=3C_2(G),\\
\label{eq:C20}
    C^{ilm}C_{jlm}&=(2g^2)C_2(R_i)\delta^i_j,
\end{align}
where $C_2(G)$ represents the quadratic Casimir operator for the adjoint representation of the group. These finiteness relations at one- and two-loops have very important implications for the matter content, as they restrict the possible choices of irreducible representations within a gauge group $G$. In particular, it is verified that under these conditions, the minimal supersymmetric standard model (MSSM) would not have the characteristic of being finite, because for the group $U(1)_Y$ it holds that $C_2[U(1)]=0$, which is inconsistent with the condition given by equation (\ref{eq:C100}). Therefore, we will work with unification models based on simple or semi-simple gauge groups of the type $SU(N)$. Likewise, this same restriction applies when considering the spontaneous breaking of supersymmetry through $F$ terms or $D$ terms, since their implementation requires the presence of a $U(1)$ gauge group, which makes them incompatible with finiteness. That is why the breaking of supersymmetry is expected to occur through the inclusion of soft terms in the Lagrangian, in order to be consistent with the finiteness conditions above stated.

The information provided to achieve finiteness at one and two loops encoded in equations (\ref{eq:C100}) and (\ref{eq:C20}), along with the search for relations invariant under the renormalization group, will form the foundation for addressing the question of how to construct a theory that is finite at all orders in perturbation theory. The latter is driven by the necessity for the relationships between parameters, such as relation (\ref{eq:C20}), to maintain their form at any renormalization scale. This is not trivial to achieve as it requires the coupling correspondences to be solutions to the reduction equations and thus solutions in power series \cite{Zimmermann:1984sx,Heinemeyer:2019vbc}. The necessary and sufficient conditions to ensure finiteness in the dimensionless sector of a supersymmetric theory at all orders are four and are condensed in the finiteness theorem, which states the following \cite{Lucchesi:1987ef,Piguet:1986pk,Lucchesi:1987he}:

\textbf{Theorem.} For a supersymmetric Yang-Mills theory $N=1$ with a simple gauge group $G$, if each of the following conditions is satisfied, 
\begin{enumerate}
    \item Free of gauge anomalies,
    
    \item The beta function of the gauge coupling is zero at one-loop level
    
    \[\beta_g^{(1)}=\frac{g^3}{16\pi^2}\Big[\sum_iT(R_i)-3C_2(G)\Big]=0,\]
    
    \item There is a solution of the form 
    \begin{equation}
        \label{eq:sol1}
        C_{ijk}=\sigma_{ijk}g,
    \end{equation}
    for the condition that renders zero the anomalous dimensions of the fields at one loop  (\ref{eq:gamma}), where $\sigma_{ijk}$ are complex numbers.
    
    \item The above solution is unique and non-degenerate when  considered as a solution to the one-loop Yukawa beta function being zero
$$   
\beta^{(1)}_{ijk}=0.$$
\end{enumerate}
    Then, the solution of the form (\ref{eq:sol1}) can be extended into a power series of the principal coupling $g$, so that the unified Yang-Mills theory will depend solely on a dimensionless parameter, whose gauge beta function is zero to all orders in perturbation theory. Consequently, this theory will be complete in the ultraviolet \cite{Lucchesi:1996ir,Heinemeyer:2019vbc}.

The detailed proof of the theorem relies on two important considerations, both related to the mathematical structure of currents and chiral anomalies in an $N=1$ supersymmetric theory \cite{Piguet:1986td,Piguet:1981mu,Piguet:1981mw,Shifman:1986zi}. To provide a comprehensible overview of the proof, a detailed analysis of the supercurrent multiplet associated with the theory is required, for which the following assumptions are made.

Consider an $N=1$ theory with a non-Abelian gauge group and possessing a component $k$ of chiral superfields in the superpotential $W(\phi_k) = \sum_a y_a W^a(\phi_l)$. This theory is endowed with invariance under chiral $R$-transformations with its respective Noether current, and also exhibits scale invariance at the classical level. However, at the quantum level, it has anomalies associated with the energy-momentum tensor. Finally, if we consider that the theory in question is massless\footnote{The massless theory exhibits scale invariance, whereas the massive theory has its scale invariance restricted by factors dependent on derivatives of the superpotential.}, it will also be classically invariant under the superconformal transformation group; which is an extension of the conformal group \cite{Cordova:2016emh},  and similarly exhibits anomalies in the quantum case.

All the information pertaining to each of the symmetries and their respective anomalous currents is contained within a supermultiplet $J$, which comprises the energy-momentum tensor associated with translational symmetry, the axial current associated with chiral $R$ invariance, and the supersymmetry current. Schematically, these can be written as follows:

\begin{equation}
    J=\{T^{\mu}_{\nu},J^{\mu}_R, Q^{\mu}_{\alpha},\dots\}.
\end{equation}

For a generic theory, this quantity is not conserved at the classical level, due to which the anti-chiral derivative of the supercurrent will be corrected by factors dependent on the superpotential. When considering this divergence at the quantum level, scale anomalous terms will arise, causing the conservation equation to also include scale-dependent factors $\mu$. Thus, the equation for the anomalies can be expressed in general form for the supermultiplet of currents as \cite{Shifman:1986zi,Leigh:1995ep}

\begin{equation}
\label{eq:FZ}
\bar{D}^{\dot{\alpha}}J_{\alpha\dot{\alpha}}=\frac{1}{3}D_{\alpha}(3W-\sum_k\phi_k\frac{\partial W}{\partial\phi_k})-
D_{\alpha}\Big[\delta TrW_{\mu}W^{\mu}+\frac{\bar{D^2}}{8}\Sigma_k\gamma_k Z_k\phi_k^{\dagger}e^V\phi_k\Big],
\end{equation}
from which the renormalization amplitudes of the superfields $Z_k$, the one-loop gauge beta function parameter $\delta = [3C_2(G) - \Sigma T(R_k)]$, and the anomalous dimension $\gamma_k$ associated with each superfield $\phi_k$, can be distinguished.
Consistent with the work of Ferrara and Zumino \cite{Ferrara:1974pz}, the above equation not only contains information about the non-conservation of the currents of the multiplet $J_{\alpha\dot{\alpha}}$, but also hints at the mathematical structure of the anomalies. According to the work described in \cite{Shifman:1986zi}, it is possible to show that the anomalies present in the current multiplet also form a new supermultiplet, with each of these anomalies having a coefficient that relates them to each other. This is noteworthy because this common coefficient is the one-loop gauge beta function, which implies that if the second condition of the finiteness theorem is satisfied, the cancellation of all anomalies within the multiplet would be achieved by $J_{\alpha\dot{\alpha}}$.

The form of the beta functions of the renormalization group for the dimensionless couplings of a theory with superpotential $W(\phi_k)=\sum_a y_a W^a(\phi_l)$, with $W^a$ being a product of $d_a$ chiral superfields, can be generically written as follows, thanks to the non-renormalization theorem
\begin{equation}
\label{eq:bnm}
    \beta_g^{(1)}=f(g)\Big[\sum_iT(R_i)-3C_2(G)-\sum_iT(\phi_i)\gamma(\phi_i)\Big]=f(g)S_g,
\end{equation}

\begin{equation}
\label{eq:byuk}
    \beta_y^{(1)}=y\Big[-d_W+\sum_i[d(\phi_i)+\frac{1}{2}\gamma(\phi_i)\Big]=yS_y,
\end{equation}
with $d_W$ and $d(\phi_i)$ being the canonical dimensions of the superpotential and the superfields, respectively. In here, $f(g)$ is a smooth function of the gauge coupling dependent on the model under consideration and varies implicitly with the renormalization energy scale. Similarly, the superpotential coupling $y_a$ varies with the scale $\mu$. The coefficients $S_g$ and $S_y$ that accompany these beta functions will encode scale invariance at a fixed point when they are zero.

Returning to the discussion around equation (\ref{eq:FZ}), according to what is described by Piguet and Sibold \cite{Piguet:1986td} and the work of Shifman and Vainshtein \cite{Shifman:1986zi}, it is possible to rewrite this conservation relation in terms of the equations of motion for the chiral superfields, as they are proportional to the terms with scale dependence

\[\phi_k\frac{\partial W}{\partial \phi_k}+\frac{T(R_k)}{16\pi^2}W_{\alpha}W^{\alpha}=\frac{\bar{D}^2}{4}Z_k
\phi_k^{\dagger}e^V\phi_k,\]
such that the anomaly can be written as
\begin{multline}
\frac{1}{3}D_{\alpha}(3W-\sum_k\phi_k\frac{\partial W}{\partial \phi_k})\\-\frac{1}{3}D_{\alpha}\Big[(\frac{\delta}{32\pi^2})W_{\beta}W^{\beta}+\frac{1}{2}\sum_k\gamma_k(\frac{1}{16\pi^2}T(R_k)W_{\beta}W^{\beta}+\phi_k\frac{\partial W}{\partial \phi_k})\Big]=\\
-\frac{1}{3}D_{\alpha}\Big[\sum_k\phi_k\frac{\partial W}{\partial \phi_k}(1+\frac{1}{2}\gamma_k)+
\frac{W_{\alpha}W^{\alpha}}{32\pi^2}\sum_k(T(R_k)\gamma_k+\delta)-3W\Big],
\end{multline}
and in terms of the factors $W^{a}$ of the superpotential it is written as follows
\begin{equation}
 \label{eq:ansc}
 \bar{D}^{\dot{\alpha}}J_{\alpha \dot{\alpha}}=-\frac{1}{3}D_{\alpha}[\frac{W_{\beta}W^{\beta}}{32\pi^2}S_g+
 \sum_k y_aW^{(a)}S_{y_a}],
\end{equation}
where $S_g$ and $S_{y_a}$ are the scaling factors. This expression makes clear the dependence of the anomaly multiplet on the parameters of the beta functions of the theory, as well as demonstrating that the quantities associated with axial anomalies, for example, are one-loop quantities only \cite{Piguet:1986td,Grisaru:1985yk}. The axial current associated with chiral $R$ transformations will be given by the $\theta = 0$ component of the superfield $J_{\alpha \dot{\alpha}}$, so that equation (\ref{eq:ansc}) for this component will represent a Ward identity for the axial current. According to the discussions in the works \cite{Shifman:1986zi,Piguet:1986td}, this current will have an associated anomaly with the following structure:
\begin{equation}
    \label{eq:anomaly}
    r=\beta_g(1+x_g)+\beta_{ijk}x^{ijk}-\gamma_Ar^A,
\end{equation}
where the quantities $x_g$ and $x_{ijk}$ are radiative quantities, and $\gamma_A$ refers to combinations of the anomalous dimensions of the matter superfields. This structure can be directly inferred from the conservation equation (\ref{eq:ansc}) where the superpotential is of the form of (\ref{eq:sup3}). Therefore, the condition of having a zero one-loop gauge beta function will be equivalent to the cancellation of the chiral $R$-anomaly. Furthermore, the vanishing of the one-loop anomalous dimension implies that the Yukawa beta functions will be zero at that order, which implies that the coefficients $r^A$ will be zero. With all this in mind, the proof of Theorem 1 is obtained \cite{Lucchesi:1987ef}.

\textbf{Proof.}

Consider the beta function for the Yukawa couplings given by the reduction equation
\begin{equation}
\label{eq:redsusy}
\beta_{ijk}=\beta_g\frac{d C_{ijk}}{dg},
\end{equation}
inserting this into the equation for the chiral anomaly (\ref{eq:anomaly}), we obtain the following
$$0=\beta_g(1+x_g)+\beta_g\frac{d C_{ijk}}{dg}x^{ijk},$$
where all the chiral anomalies are zero, same as the coefficients $r^A$ \cite{Piguet:1986td}. The homogeneous equation for $\beta_g$ can be rewritten as
$$0=\beta_g(1+O(\hbar)),$$
which has a solution in the perturbative sense in powers of $\hbar$, with $\beta_g=0$ order by order. This used the fact that the coefficients $x_g$ and $x_{ijk}$ are one-loop quantities, or in other words, are at order $\hbar$. Therefore, the theory will be ultraviolet complete thanks to the reduction equation, since $\beta_{ijk}=0$ to all perturbative orders. $\blacksquare$

There are interesting consequences to highlight from this theorem at this point. The first of these is that anomalies, by sharing parameters in the supermultiplet, will cancel out to all orders in perturbation theory, resulting in an anomaly-free theory in a highly symmetric regime. Particularly, without anomalies present in the theory at the Grand Unification scale, scale invariance will be maintained, and the beta functions at the fixed point will vanish. This results in a relationship between the couplings, such that if there are \( k \) couplings in a theory, there will be the same number of conditions. The solution of the reduction equations at the fixed point determines a submanifold in the coupling space, whose dimension is determined by the number of linearly independent solutions, as previously mentioned.

If the assumption is made that there exists a fixed point such that the theory is conformal (superconformal), the directions of the flow will be determined by a set of operators whose addition to the potential will generate new fixed points. These operators, called marginal operators, generate a differentiable manifold of fixed points\footnote{Do not confuse the solution manifold at a fixed point with the manifold of fixed points.} or a conformal manifold associated with a CFT. In the context of superconformal theories, the existence of such operators stands out in limited cases, depending on the spacetime dimensionality $d$ and the number of supermultiplets present, being viable only for $N=1,2,4$ in $d=4$ and $N=1,2$ in $d=3$ \cite{Cordova:2016emh}, where a characterization of the unitary representations of the superconformal algebra is given, restricting the scaling dimension and $R$ charges according to their recombination rules for operators.

One of the reasons why the assumption of supersymmetric conformality at the fixed point is so powerful is that such theories exhibit the conservation of chiral $R$-symmetry at the quantum level. In other words, the algebra of continuous $R$ transformations is contained within the superconformal algebra. Thus, if there exists a superconformal fixed point, the presence of $R$ anomalies will be suppressed, and the operators in the theory will form a representation of the superconformal group. If it is consistently defined that the multiplicative factors accompanying the beta functions (\ref{eq:bnm}) and (\ref{eq:byuk}) properly encode the scale dependence in the running of parameters, then for a Superconfromal Field Theory (SCFT), these scale coefficients $S_y$ and $S_g$, which we will call scaling coefficients, must necessarily vanish. Therefore, this discussion of conformality is strongly connected to having a finite theory at all orders. The next section will examine in detail how the invariance under the superconformal group of the theory is a direct consequence of the finiteness conditions given by Theorem 1 in certain cases.

\section{Relation between Finiteness and Conformal Invariance}

The finiteness conditions are a significant outcome in supersymmetric theories, as their absence of divergences in the effective action makes them particularly attractive for developing high-energy unification models as extensions of the Standard Model that are also phenomenologically viable \cite{Heinemeyer:2014vxa,Heinemeyer:2019vbc}. However, it is likely that the conditions of Finiteness Theorem  are more powerful than initially thought, as the theory being scale-invariant at a fixed point $\mu^{*}=M_{GUT}$ would set the stage for the existence of a conformal manifold associated with that fixed point, thus creating a model with a much higher degree of symmetry, linked to each zero of the beta function. Conformal manifolds, according to Conformal Field Theory (CFT) \cite{Blumenhagen:2009zz,Gukov:2015qea,Behan:2017mwi}, are generated by marginal operators linked to directions within the fixed manifold, thereby characterizing its dimensionality within parameter space in the Wilsonian sense. The presence of such operators in a quantum field theory implies the existence of a supersymmetric conformal manifold formed by operators acting in superspace, provided the theory remains SUSY invariant at the fixed point (in fact, the effective action will exhibit invariance under super-Weyl transformations).

Taking into account the algebraic structure of superconformal theories \cite{Cordova:2016emh}, the following theorem/proposition is stated regarding symmetry within Finite Grand Unification Theories.

\textbf{Proposition.} Let there be a supersymmetric $N=1$ gauge theory in $d=4$, with gauge group $G$ and superpotential

\[
W = m_{ij}\Phi_i\Phi_j + C_{ijk}\Phi_i\Phi_j\Phi_k,
\]
such that conditions 1-4 of Theorem are satisfied. Then, if a fixed point associated with an energy scale $\mu^{*}$ exists, the theory is invariant under the superconformal group at that point.

As a consequence, if this theorem holds, the operators of the supersymmetric theory will constitute a unitary representation of the superconformal group in $d=4$, as well as the current multiplet corresponding to the symmetries of the theory. Additionally, the conformal theory will include in its manifold sets of coefficients associated with the primary operators, these are known as the coefficients of the Operator Product Expansion (OPE) and scaling dimensions or conformal dimensions $\lambda_{ijk}$ and $\Delta_i$,  respectively. The condition of marginality in the operators will be reflected in conditions on these coefficients, such that \cite{Behan:2017mwi}

$$
\centering
\begin{array}{cc}
 \hat{\Delta}=d,&\lambda_{\hat{\mathcal{O}}\hat{\mathcal{O}}\hat{\mathcal{O}}}=0,
 \end{array}
$$

Constraint that will result in a fundamental tool for the search for fixed manifolds. \\
\textbf{Proof.} \\
A superconformal theory must satisfy three fundamental conditions in terms of its operators and its symmetry group, these conditions arise from the algebraic study of SCFTs and from what has been described by various authors \cite{Leigh:1995ep,Cordova:2016emh,Behan:2017mwi}:

\begin{enumerate}
 \item Scale invariance must be verified, encoded in the scaling coefficients $S_g$ and $S_{C_{ijk}}$.
 \item The chiral $R$ current must be conserved at the quantum level, so the $\theta=0$ component within the multiplet $J_{\alpha \dot{\alpha}}$ must be zero.
 \item There must exist marginal operators\footnote{Local primary operators whose anomalous dimension is zero.} that will induce the theory to manifest in a superconformal regime.
\end{enumerate}

Since we are interested in a finite theory at the Grand Unification scale $M_{GUT}$, such that the reduction of couplings takes place and therefore solutions to equation (\ref{eq:sol1}) exist, SCFT invariance is sought at the fixed points of the theory.

The first point is trivially verified for the superpotential according to conditions 1 and 3 of the finiteness theorem, as the scale coefficients are proportional to these conditions around $\mu^{*}=M_{GUT}$. Point number 2 can be verified thanks to the cancellation of anomalies within the current supermultiplet as a consequence of finiteness (\ref{eq:anomaly}), in particular, the chiral $R$ symmetry is conserved at the quantum level \cite{Piguet:1986td,Heinemeyer:2019vbc}.

The third condition turns out to be the most subtle to prove, as the way to show that there are exactly marginal operators and, therefore, a superconformal fixed manifold, is through the direct exhibition of the operators. For this, an operator is proposed as a candidate for marginality, suggesting the superpotential term whose coupling influences the critical properties of the theory $\mathcal{O}_1=C_{ijk}\phi_i\phi_j\phi_k$. According to the characterization given in \cite{Leigh:1995ep}, this operator verifies that its scaling factor $S_{C}$ is proportional to the scaling factor of the gauge coupling. This is due to condition 2 of the finiteness theorem and the non-renormalization theorem, which provide a particular form for the Yukawa beta function proportional to the anomalous dimensions $\gamma_{i}^{j}$, obtaining that the gauge scale coefficient is

\begin{equation}
    S_g=\Sigma_kT(R_k)\gamma_k,
\end{equation}
with $\gamma_k$ the diagonal matrix elements of the anomalous dimension. 
Therefore, the Yukawa scaling factor is given by
\begin{equation}
    S_C=\frac{3}{2}\gamma_k.
\end{equation}
With this, it is proven that the operator $\mathcal{O}_1$ is a good candidate to be a marginal operator since $S_g \propto S_C$ \footnote{For all values of $i$, $j$, and $k$.}. This condition can also be deduced by taking a solution to the reduction equation (\ref{eq:redsusy}) given by Theorem 1, so that this relationship between $S_g$ and $S_C$ will be verified order by order, and it makes it clear that only the trilinear terms that can be marginal are those that satisfy the finiteness condition. Another aspect of $\mathcal{O}_1$ is the absence of anomalous dimension associated with each of the superfields that compose it, which occurs because, according to condition 3 of Theorem 1, these vanish for all values of $i$ and $j$.

Finally, since we are dealing with chiral operators in a $d=4$ dimension theory, according to \cite{Cordova:2016emh} there exists a close relationship between the canonical dimension of the field and its charge under the $R$ symmetry, which is given by $d_k = \frac{3}{2}r_k$. As already mentioned, the condition for marginal operators is that their canonical dimension matches their anomalous dimension; that is, $d_k = \Delta_k$, meaning that $\Delta_k = \frac{3}{2}r_k$. This makes the operators of the finite theory good candidates to reside in a conformal supermultiplet, according to the explicit classification given in ref.~\cite{Cordova:2016emh}. In particular, since the operators that constitute the marginal ones have vanishing anomalous dimension $\gamma_i^j$, they will satisfy the relationship between $r_k$ and $\Delta_k$, since in terms of the anomalous dimension, the relationship is written as $\frac{3}{2}r_k = 1 + \frac{1}{2}\gamma_k$. 

Therefore, $\mathcal{O}_1$ has sufficient elements to reside in a conformal multiplet as previously discussed, and it is concluded that it is a marginal operator in the finite theory at the fixed point $\mu^{*} = M_{GUT}$.
Consequently, this demonstrates that a conformal manifold emerges at the Grand Unification scale, characterized by a  finite effective action of the  dimensionless-sector and the operators can  assemble into a unitary representation of the superconformal group. 

It is worth highlighting that, as with the conditions for finiteness to all orders in perturbation theory, the assumption of reduction of couplings in the theory at the Grand Unification scale is crucial when proving the previously stated conformal theorem. This emphasizes that employing a reduction in the dimensionless couplings of the theory reduces the parameter space in such a way that results in an associated conformal symmetry, and therefore possible extra relations between the parameters will not depend solely on the addition or existence of marginal operators, it is also linked to the RGI (Renormalization Group Invariant) relations between the parameters, which hold at all orders in perturbation theory. As it will be shown, the conditions of reduction of couplings and finiteness remain compatible even with supersymmetry breaking (with some exceptions) within the model, sacrificing the conformal invariance in the process but not the effect that this symmetry has on the couplings. 

\section{Reduction of Couplings and SUSY breaking}

The idea of constructing a finite supersymmetric Grand Unified Theory goes hand in hand with the goal of reducing the number of free parameters in the Standard Model. For such models to be phenomenologically consistent with experimental observations at low energies, it is necessary to include the mass sector associated with supersymmetry breaking. Since this sector has a characteristic scale $m_0$, it is expected to break conformal symmetry in the dimensionless sector. To test this assumption, one uses the Zimmermann’s method for dealing with parameters of dimension 1 and 2, the latter appearing in the soft breaking terms.

Let us consider an $N=1$ supersymmetric gauge theory based on a gauge group $G$, with a superpotential identical to equation (\ref{eq:sup3}), such that
\begin{equation}
    W=\frac{1}{2}M^{ij}\Phi_i\Phi_j+\frac{1}{6}C^{ijk}\Phi_i\Phi_j\Phi_k,
\end{equation}
with $\Phi_i$ chiral superfields. It will also be assumed that the supersymmetry breaking gives rise to soft breaking terms identical in form to those studied for the Minimal Supersymmetric Standard Model (MSSM), which are constrained in form by phenomenological considerations and are given by
\begin{equation}
    -\mathcal{L}_{SSB}=\frac{1}{6}h^{ijk}\phi_j\phi_j\phi_k++\frac{1}{2}b^{ij}\phi_i\phi_j+\frac{1}{2}(m^2)^i_j\phi^{*j}\phi_i+\frac{1}{2}M\lambda\lambda+h.c.,
\end{equation}
where factors responsible for baryon number ($B$) and lepton number ($L$) violation were not considered. In turn, one identifies dimension-one parameters, $h^{ijk}$, and dimension-two parameters $b^{ij}$, $(m^2)^i_j$, and $M$\footnote{Not to be confused with the $M^{ij}$ parameters in the superpotential.}. The latter corresponds to the gaugino mass and generally carries an index $a$, referring to the number of gauginos present depending on the structure of the gauge group in the theory.
     
The renormalization group functions $\beta^{(1)}_g$, $\beta^{(1)}_{ijk}$ and $\gamma^{(1)i}_j$ preserve the same form as in equations (\ref{eq:bet}-\ref{eq:gamma}). We will also assume that it is possible to perform the reduction of the dimensionless parameters $C^{ijk}$ and $g$, meaning that there exists a power series solution to the reduction equations of the form
\begin{equation}
    \label{eq:solutionRGI}
    C^{ijk}=\sum_{n=0}\sigma_{(n)}^{ijk}g^{2n+1}.    
\end{equation}

Using the all-loop relations among the soft breaking terms (SSB) \cite{Hisano:1997ua,Jack:1997eh,Jack:1999aj}, derived by means of the spurion technique \cite{Fujikawa:1974ay,Grisaru:1979wc}, and the fact that the following relation is RGI invariant \cite{Jack:1997eh}
\begin{equation}
\label{eq:softred}
h^{ijk}=-M\frac{dC(g)^{ijk}}{d\ln{g}},
\end{equation}
—i.e., that it is possible to relate the couplings $h_{ijk}$ and $C_{ijk}$ to all orders in perturbation theory—one can derive the following renormalization group invariant relations \cite{Hisano:1997ua,Jack:1999aj}
\begin{align}
\label{eq:masg}
    M&=M_0\frac{\beta_g}{g} ,& h^{ijk}=&-M_0\beta_{C^{ijk}},\\
    \label{eq:b2}
    b^{ij}&=-M_0\beta^{ik}_M ,&  (m^2)^i_j=&\frac{1}{2}|M_0|^2\mu\frac{\partial \gamma^i_j}{\partial \mu},
\end{align}
where $M_0$ is a reference scale to be specified. Taking into account the above together with the finiteness theorem conditions, it is possible to replace the relation involving the scalar masses in (\ref{eq:b2}) with a more general one, a sum rule that governs the behavior of scalar masses to all orders in perturbation theory and that retains its form across different energy scales \cite{Kobayashi:1997qx,Kobayashi:1998jq}
\begin{align}
    \label{eq:sumrule}
    m_i^2+m_j^2+m_k^2=|M|^2\Big\{&\frac{1}{1-g^2C_2(G)/(8\pi^2)}\frac{d\ln{C^{ijk}}}{d\ln{g}}+\frac{1}{2}\frac{d^2\ln{C^{ijk}}}{d(\ln{g})^2}\Big\}\,+\notag \\&\sum_l\frac{m_l^2T(R_l)}{C_2(G)-8\pi^2/g^2}\frac{d\ln{C^{ijk}}}{d\ln{g}} ~.
\end{align}

For the derivation of this sum rule, the evolution equation for the mass parameters $m_i^2$ was used, together with the constraints given by the finiteness theorem and the explicit form of the gauge beta function in the NSVZ scheme, given by \cite{Kobayashi:1998jq}
\begin{equation}
    \label{eq:NSVZ}
    \beta^{NSVZ}_g=\frac{g^2}{16\pi^2}\Bigg[\frac{\sum_{l}T(R_l)(1-\gamma_l/2)-3C_2(G)}{1-g^2C_2(G)/8\pi^2}\Bigg],
\end{equation}
this expression is valid to all orders in perturbation theory for $N=1$ supersymmetric Yang–Mills theories and is obtained by considering an instanton background in the model. According to instanton theory \cite{Vandoren:2008xg}, it is necessary to take into account the contributions of both bosonic and fermionic zero and non-zero modes in the integration measure of the effective action. As a result, the contributions from non-zero modes, as well as possible non-perturbative effects, cancel out at higher orders in the loop expansion. This last condition is very important, as it ensures that equations (\ref{eq:masg}) and (\ref{eq:b2}) will not receive non-perturbative contributions.

Assuming a specific power series solution of the
reduction equations of the form of Eq.~(\ref{powerser}) to the first order in $g$, equation (\ref{eq:sumrule}) reduces to the following form 
\begin{equation}
    \label{eq:sumrule2}
    m_i^2+m_j^2+m_k^2=|M|^2,
\end{equation}
which is model independent \cite{Heinemeyer:2019vbc}\footnote{In the finite models studied so far, the second-order contribution to the sum rule turns out to be zero.}, and depends solely on a single parameter, $|M|^2$. This type of sum rule arises in various scenarios and theories, particularly in certain classes of superstring models \cite{Brignole:1995fb,Ibanez:1998xn}, and also appears to be universal in models with gauge–Yukawa unification based on the reduction of couplings method \cite{kawamura1997soft,Kobayashi:1997qx}. Upon closely examining the relations obtained for the parameters of the soft-breaking Lagrangian, one observes that they are identical to those found in the literature for a scenario in which soft terms are generated via supersymmetry breaking mediated by the violation of conformal symmetry \cite{Randall:1998uk,Gherghetta:1999sw, Giudice:1998xp}, particularly if the reference scale $M_0$ is taken to be the gravitino mass $m_{3/2}$.

The specific form of such expressions suggests that superconformal invariance in the Grand Unified theory must be broken in some limit, most likely simultaneously with the supersymmetry breaking, thus generating this type of invariant terms. However, the appearance of these terms will also depend on the suppression of tree-level gravitational effects. Therefore, any assumption made in order to obtain the relations (\ref{eq:masg}) and (\ref{eq:b2}) is expected to be tied to the suppression of such gravitational effects, at least at tree level in the perturbative expansion \cite{Chung:2003fi}.

Furthermore, the existence of a sum rule for scalar masses is highly useful for studying the phenomenology of finite models. As mentioned already above, the set of relations (\ref{eq:masg}) and (\ref{eq:b2}) coincides with the soft supersymmetry-breaking terms of the type generated through conformal anomaly mediation \cite{Jack:1999aj}, by fixing $M_0$ to be $m_{3/2}$. 
The key difference in having the sum rule as a boundary condition for the RGEs at the unification scales, is that it allows to have non-tachyonic slepton spectra at low energies, thus rendering phenomenologically viable models \cite{Kobayashi:1997qx,Heinemeyer:2019vbc}. 
Moreover, since the couplings in the soft sector are invariant under the renormalization group, they are blind to flavor changes. As a result, the phenomenological constraints usually imposed on supersymmetry-breaking models, i.e. those typical of the MSSM, are satisfied in finite models with this type of soft breaking terms.

The goal at this point is to provide a satisfactory answer to the questions that arise when working with the soft sector in the context of reduction of couplings. In particular, we aim to understand the coincidence between relations (\ref{eq:masg}) and (\ref{eq:b2}) with those appearing in the scenario of anomaly-mediated supersymmetry breaking (AMSB). Guided by the claimed correspondence between flat-space conformal theories and curved-space Weyl-invariant theories \cite{Farnsworth:2017tbz}, we proceed to evaluate the effects of Zimmermann’s reduction process on a super-Weyl invariant framework. This methodology is strongly motivated by the equivalence between finiteness and conformal invariance in Finite Grand Unified Theories, as well as the fundamental idea that supergravity provides the necessary structural connection for a divergence-free supersymmetric model at high energy scales \cite{Nanopoulos:1994as}.

Let us now assume that our finite theory before supersymmetry breaking corresponds to a low-energy effective theory of an $N=1$ supergravity theory. In this setup, the superfields associated with the latter will be responsible for mediating SUSY breaking through the communication between the hidden sector and the visible sector, where the fields of the visible sector are denoted by $\Phi_i$ and the fields of the hidden sector by $X$.

To ensure that the dominant contribution to the soft parameters after SUSY breaking is the one corresponding to the conformal anomaly, it is required that the potential contributions from gravity are suppressed at tree level. This is achieved by first constructing a scale-invariant action, thereby eliminating possible direct couplings between the $X$ and $\Phi$ fields in the superpotential $W$ and the Kähler potential $K$ \cite{kaplunovsky1994field}. For this purpose, let us consider the most general Lorentz-invariant SUGRA Lagrangian in $d=4$
\begin{equation}
\label{eq:Lsugra}
    \mathcal{L}=\int d^4\theta\, E\,\mathcal{F}(\bar{\Phi},e^{-V}\Phi)+\int d^2\theta\, \varepsilon (W(\Phi)+f(\Phi)\mathcal{W}^2_{\alpha})+h.c.,
\end{equation}
where $E$ and $\varepsilon$ are the Vielbein superfields, $W$ is the superpotential, $f$ is the  complex  gauge-kinetic function, $\mathcal{W}_{\alpha}$ the field-strength chiral superfield, and $\mathcal{F}$ is a continuous function depending on the supergravity Kähler potential. This expression is related to the usual supergravity Lagrangian \cite{ellis1987supersymmetry} through a Weyl transformation, or Weyl rescaling of the spacetime metric
\begin{equation}
    \label{eq:WeylTrans}
    g_{\mu\nu}\rightarrow g_{\mu\nu}e^{-\frac{1}{3}K/M_{pl}^2},
\end{equation}
and its purpose is to ensure that the bosons and fermions in a supermultiplet are normalized in the same way.

This Lagrangian by itself does not exhibit scale invariance or, more generally Weyl invariance. However, it is possible to construct a SUGRA Lagrangian that is Weyl invariant and at the same time ensures that the associated rescaling is supersymmetric. Transformations of this kind are known as Super-Weyl transformations \cite{kaplunovsky1994field} and are parametrized by a chiral superfield $\tau$, such that the chiral fermions of the model transform as\footnote{The complete list for all the superfield transformations can be reviewed in \cite{kaplunovsky1994field}}
\begin{align}
\label{eq:ChiralTrans}
    \lambda_{\alpha}^{(a)}\rightarrow \lambda_{\alpha}^{(a)} \cdot e^{-3\tau}\\
    \psi_{\alpha}^{i}\rightarrow \psi_{\alpha}^{i} \cdot e^{\tau-2\bar{\tau}}.
\end{align}

In this way, the associated supergravity theory can be symmetrized through the introduction of a new non-dynamical chiral superfield $\varphi$, which will transform under super-Weyl transformations as
\begin{equation}
    \label{eq:Compensator}
    \varphi\rightarrow e^{-2\tau}\cdot \varphi,
\end{equation}
such that, by introducing this field multiplicatively into the Lagrangian with an appropriate number of powers of $\varphi$ and $\bar{\varphi}$, the super-Weyl rescaling will be \textit{compensated}, yielding a Weyl-invariant Lagrangian, this formalism is known as the Weyl compensator.
The immediate effect of introducing the compensator into the SUGRA action is that the expressions for the supersymmetric potentials $K$ and $W$, as well as the gauge kinetic function $f$, will be modified in form and will now depend on this new superfield. The new form is the following
\begin{align}
    \label{eq:phidependence}
    &\tilde{K}(\varphi,\bar{\varphi},\Phi, \bar{\Phi})=K-6M_{\text{pl}}^2(Re\ln{\varphi}) ~,\\
    &\tilde{W}(\varphi,\Phi)=\varphi^3 \,W(\Phi) ~,\\
    &\tilde{f}(\Phi)=\varphi^0  \, f(\Phi)~.
\end{align}

The expression of the super-Weyl invariant Lagrangian then reduces to the form
\begin{equation}
    \label{eq:LsuperWeyl}
    \mathcal{L}=\int d^4\theta\, \mathcal{F}(\Phi,\bar{\Phi})\,\varphi\bar{\varphi}+\int d^2\theta\, (\varphi^3\, W(\Phi)+f(\Phi)\,\mathcal{W}_{\alpha}^2)+ h.c.,
\end{equation}
    where it is very explicit that the compensator plays the rôle of making the Weyl invariance manifest, as well as breaking it when it acquires a non-zero vacuum expectation value $\langle \varphi \rangle\neq 0$. In this last expression, the identification of the gravitational fields $E$ and $\varepsilon$ was made so that the compensator superfield $\varphi$ contains all their information \cite{gates2001superspace}, with the auxiliary component of $\varphi$ being precisely the spin-zero component of the SUGRA field. In other words, $\varphi = 1 + F_{\varphi}\theta^2$, which can be interpreted as a gauge-fixing in superspace for the gravitational fields \cite{Cheung:2011jp}. 

The expression in (\ref{eq:LsuperWeyl}), although scale (Weyl) invariant, by itself is not sufficient to eliminate possible interactions between the hidden and visible sectors at tree level, since the term $\mathcal{F}(K(\Phi,\bar{\Phi}))$, as a function of the Kähler potential, can still induce coupling terms between the fields $X$ and $\Phi$ through curvature factors \cite{Cheung:2011jp}. Therefore, to ensure a complete decoupling between the sectors via the supergravity field, it is necessary to impose that their separation be total (at least in this limit before supersymmetry breaking). That is, analogously to the discussion by Randall and Sundrum \cite{Randall:1998uk}, this separation must be made explicit in terms of the potentials and the scheme function, as follows:
\begin{align}
    \label{eq:DecoupleSectors}
    &\mathcal{F}=\mathcal{F}_{vis}+\mathcal{F}_{hid},\\
    &W=W_{vis}+W_{hid},
\end{align}
thus, imposing this type of structure on $\mathcal{F}$ will significantly affect the form that the Kähler potential must take in such models.

To derive the explicit form of the Kähler potential with decoupled matter sectors, we analyze the limiting case in which gravitational interactions are not relevant, that is, when $g_{\mu\nu} = \eta_{\mu\nu}$ and $\varphi = 1$. In this way, we can determine the conditions on $K$ and $\mathcal{F}$ required to ensure that the fields $X$ and $\Phi$ do not mix, and that their only tree-level interaction is their common coupling to the supergravity compensator field $\varphi$.

We should also note that no additional conditions are needed for the superpotential, since by the non-renormalization theorems \cite{salam:1975nr}, the fields in $W_{vis}$ and $W_{hid}$ will not mix. Therefore, the Weyl-invariant expression for this term will suffice for our purposes.

In the case of the Kähler potential, it is not possible to simply consider the sum of $K_{hid}$ and $K_{vis}$ in the zero-gravity limit, since all possible higher-order contributions that mix the fields from each sector must be taken into account.
Taking this into consideration and omitting possible contributions involving derivative terms of the fields ($D^2\Phi$ and  $\bar{D}^2\Phi$), the expansion is carried out in combinations that preserve scale invariance at higher orders, that is, even products of the fields $\Phi$ and $X$. In this way, one obtains that the Kähler potential satisfying condition (\ref{eq:DecoupleSectors}) which ensures the complete separation of the sectors and therefore suppresses tree-level coupling terms is given by the following expression:

\begin{equation}
\label{eq:K22}
    K=-\bar{M}^2\ln\Big({1-\frac{f(\Phi,X)}{\bar{M}^2}}\Big),
\end{equation}
where $\bar{M}$ is a high reference scale, and the function $f(\Phi, X)$ is a function of both the hidden and visible sector fields, which can be identified with the scheme function $\mathcal{F}$. 

This potential then couples to the supergravity fields when gravitational interactions become relevant. With this expression, we ensure that the Lagrangian (and therefore the action) (\ref{eq:LsuperWeyl}) possesses Weyl invariance and that no tree-level terms arising from gravity are induced that would break supersymmetry. Additionally, this kind of Kähler potential exhibits a peculiarity, since the associated kinetic terms are invariant under the non-compact group of isometry $SU(n,1)$ \cite{Ellis:1983ew,Ellis:1984bm}, where $n$ is the number of matter supermultiplets in the hidden and visible sectors. Thus, expression (\ref{eq:K22}) is equivalent to some type of Kähler potentials used in the context of no-scale supergravity in $d=4$, for the particular case in which no gauge singlets are present \cite{Ema:2024sit}.

Now that we have a complete description of a conformally (Weyl) invariant theory in which the form of the soft parameters is fully determined by the conformal anomaly, we will analyze what happens to the parameter space if at the same time one assumes the reduction of couplings in the dimensional sector of supersymmetry breaking. To do this, we will examine the generation of the soft-breaking terms in this model at the level of the renormalization behavior of physical quantities in superspace. This is usually studied in different types of schemes, such as the already-mentioned NSVZ scheme, the dimensional reduction (DR) scheme, and even the so-called holomorphic regularization schemes \cite{Boyda:2001nh}, the latter being our choice due to the clarity with which renormalization amplitudes can be handled. 
In this scheme, the information about supersymmetry breaking is entirely tied to the form of the renormalization amplitudes of the couplings and matter fields $\mathcal{Z}_0$.

According to the ref.~\cite{kaplunovsky1994field}, in the Wilsonian language, if one has a UV-regulated theory such that the cutoff $\Lambda$ preserves supersymmetry, then in order for the bare effective action and $\Lambda$ to be manifestly invariant under local transformations, particularly Weyl transformations, it is necessary for the latter to depend on the conformal compensator ($\varphi$-dependent), that is

\begin{equation}
    \Lambda_W\equiv\Lambda(\varphi),
\end{equation}
in such a way that the renormalization wavefunction $\mathcal{Z}_0$ of the matter superfields will have an implicit dependence on $\varphi$

\begin{equation}
\label{eq:Zlambda}
    \mathcal{Z}_0=\mathcal{Z}_0(\Lambda(\varphi)). 
\end{equation}

Thus, the Kähler term in the Wilsonian action takes the form
\begin{equation}
    \mathcal{L}=\int d^4\theta \mathcal{Z}_0(\Lambda_W)K(\Phi, \bar{\Phi}),
\end{equation}
where, since we are working in the holomorphic regularization scheme, the renormalization wavefunction factor is promoted to a superfield in superspace, with dependence only on the Grassmann coordinates $\theta$ and $\bar{\theta}$, and not on the spacetime coordinates \cite{Arkani-Hamed:1998mzz}.

Through the non-renormalization theorems of the Wilsonian superpotential, we know that the possible divergences associated with the matter sector must arise from the wavefunction renormalization $\mathcal{Z}$. Therefore, upon regularizing the theory, one also expects supersymmetry to be preserved and that the divergences that appear order by order be absorbed by supersymmetric counterterms. That is, the bare amplitude takes the form

\begin{equation}
    \mathcal{Z}_0=\mathcal{Z}(\mu)+\delta \mathcal{Z}(g, C^{ijk}, \Lambda_W/\mu),
\end{equation}
where $\delta \mathcal{Z}$ is the counterterm function, which depends directly on the dimensionless parameters of the superpotential and on the gauge coupling, in addition to its aforementioned dependence on the conformal compensator $\varphi$.
As in non-supersymmetric quantum field theories, these counterterms can be defined after performing an analysis of supergraph Feynman diagrams for vertices and propagators with renormalized couplings, and choosing them so as to cancel the associated divergences \cite{Grisaru:1984qj}. In this way, the dependence of the renormalization amplitude on the dimensionless parameters is the same as that of the counterterms, and the relation between the bare and renormalized quantities is given by the following equation

\begin{equation}
    \mathcal{Z}_0=\mathcal{Z}(\mu)[1+C(g,C^{ijk},\Lambda_W/\mu)],
\end{equation}
where $\mathcal{Z}(\mu)$ contains the information about the renormalization scale, and the function $C$ encodes the dependence on the dimensionless couplings and on the Weyl compensator.

Let us additionally recall that we are looking for fixed points in the parameter space for the soft-breaking couplings $h^{ijk}$, that is, renormalization group invariant relations (RGI) involving the dimensionless parameters, namely those satisfying equation (\ref{eq:softred}). In the Wilsonian language, $C^{ijk}$ has a dependence on the compensator

\begin{equation}
    C^{ijk}(\varphi,g)=\alpha(\varphi)C^{ijk}(g),
\end{equation}
with $\alpha(\varphi)$ being a smooth function of $\varphi$. Therefore, the soft trilinear breaking parameters will also depend on the conformal compensator, provided that a solution of the form (\ref{eq:solutionRGI}) to the Yukawa coupling reduction condition exists. In this way, one can infer that the dependence of the soft Wilsonian coupling $h^{ijk}$ will be of the following form,

\begin{equation}
    \bar{h}^{ijk}\equiv \frac{h^{ijk}}{\alpha(\varphi)}\propto C^{ijk},
\end{equation}
leaving the wave-function renormalization with a direct dependence on this new dimensionless coupling

\begin{equation}
\label{eq:zetacount}
    \mathcal{Z}_0=Z\Big[1+C(g,\bar{h}^{ijk},\frac{\Lambda_W}{\mu})\Big].
\end{equation}

Now, since we are interested in the evolution of the parameters as a function of energy, we need to determine the renormalization group functions, or characteristic velocities of the renormalization flow, which are given in terms of derivatives of the quantity $\ln \mathcal{Z}_0$. To this goal, we will write the previous expression for $\mathcal{Z}$ as the following logarithmic expansion

\begin{equation}
\label{eq:logZ}
    \ln{\mathcal{Z}_0}=\ln{Z}+\sum_k^\infty c_k \ln^k\Big(\frac{\Lambda(\varphi)}{\mu}\Big),
\end{equation}
where the coefficients of the series $c_k$ correspond to functions of the dimensionless parameters, in particular of $\bar{h}^{ijk}$. Since super-Weyl invariance is assumed from the outset, the cutoff $\Lambda$ will appear only in combination with the regulator as $\Lambda \cdot\varphi = \Lambda_W$. This setup is possible in scenarios with Pauli–Villars regulators or even with higher-derivative regularization \cite{Arkani-Hamed:1997qui, kaplunovsky1994field, Warr:1986we}, and moreover it satisfies the essential requirement of preserving supersymmetry.

If we now expand Eq. (\ref{eq:logZ}) in a power series of the Grassmann variables up to order $\theta^2 \bar{\theta}^2$ with respect to the compensator $\varphi$, we obtain the following relation

\begin{align}
    \ln{\mathcal{Z}_0}&=\ln{Z}+(\varphi\,\theta^2+\varphi^{\dagger}\,\bar{\theta}^2)\sum_kc_kk\ln^{k-1}\Big(\frac{\Lambda}{\mu}\Big)+(\varphi^2\theta^2\bar{\theta}^2)\sum_kc_kk(k-1)\ln^{k-2}\Big(\frac{\Lambda}{\mu}\Big),\\
    \label{eq:logZ0exp}
    \ln{Z}_0|_{\theta^2(\bar{\theta}^2)}&\simeq(\varphi\,\theta^2\,\bar{h}^{ijk}(\varphi) \,+\,\varphi^{\dagger}\,\bar{\theta}^2\,\bar{h}^{ijk}(\varphi^{\dagger}))\Big[\ln{\Big(\frac{\Lambda}{\mu}\Big)}\Big]=   \Big(\varphi\,\theta^2\frac{h^{ijk}}{\alpha(\varphi)}\,+\,\varphi^{\dagger}\,\bar{\theta}^2\frac{h^{ijk*}}{\alpha(\varphi^{\dagger})}\Big)\Big[\ln\Big(\frac{\Lambda}{\mu}\Big)\Big],
\end{align}
that, to linear order in the trilinear coupling, the wave-function renormalization has a scale dependence of the following form
\begin{equation}
    \ln{\mathcal{Z}_0}|_{\theta^2(\bar{\theta}^2)}\simeq\,(h^{ijk}\,\theta^2+h^{ijk*}\,\bar{\theta}^2)\Big[\ln\Big(\frac{\Lambda}{\mu}\Big)\Big].
\end{equation}

In this last expression, two very interesting features can be observed. First, the dependence of $\ln Z_0$ on the conformal compensator has disappeared explicitly, due to its dependence on the Wilsonian coupling $\bar{h}^{ijk}$. This cancellation relies entirely on the existence of a solution to the  reduction of couplings condition for the dimensionless parameters, as well as on the reduction condition in the soft sector (\ref{eq:softred}).
The second observation follows directly from the first: since there is no explicit dependence on $\varphi$, the information about the renormalization of the theory, encoded in $\mathcal{Z}_0$, depends on a compensator-independent cutoff. As a result, we are working in a scheme in which $\Lambda$ still preserves supersymmetry, but, being $\varphi$-independent, no longer respects Weyl invariance.

We can therefore conclude from this analysis that the process of reducing dimensional parameters in the soft sector, within a highly symmetric theory, induces the breaking of conformal symmetry (through the anomaly), and consequently makes the effects of the conformal anomaly dominant over any other contribution to supersymmetry breaking.

Let us consider once again the expression (\ref{eq:zetacount}). If we now perform a Grassmann expansion in powers with respect to the energy-scale dependence, and using that $\gamma_i = \mu\, d\ln Z / d\ln \mu$, we obtain the following expression

\begin{equation}
\label{eq:zetamu}
    \ln{\mathcal{Z}}_0=\ln{Z}+\theta^2\frac{d\,\ln{Z}}{d\,\ln{\mu}}\,F_{\varphi}-\frac{1}{2}\mu\frac{d\,\gamma_i}{d\,\mu}\,F_{\varphi}^2\,\theta^2\,\bar{\theta}^2\,+\,h.c.,
\end{equation}
where, once conformal symmetry is broken, one has $\langle \varphi \rangle \simeq F_{\varphi}$. By identifying the terms in the expansions of Eqs. (\ref{eq:logZ0exp}) and (\ref{eq:zetamu}), one finds\footnote{In making this identification, we have used the fact that the $\theta^2 \bar{\theta}^2$ component of the wave-function renormalization is proportional to the soft scalar squared masses \cite{Boyda:2001nh, Arkani-Hamed:1998mzz}.}

\begin{align}
\label{eq:h&m}
    -\gamma_{ijk}F_{\varphi}\propto h_{ijk},&&\frac{1}{2}\mu\frac{d\,\gamma_i}{d\,\mu}F_{\varphi}^2\propto m_i^2,
\end{align}
where $\gamma_{ijk}$ is a linear combination of the anomalous dimensions of the superfields. These expressions are quite similar to the relations previously obtained under the idea of constructing renormalization group invariants (RGI) in Eqs. (\ref{eq:masg})–(\ref{eq:b2}) \cite{Hisano:1997ua}, upon identifying $F_{\varphi}$ with $M_0$.

A similar analysis can be carried out for the gaugino mass term, since in the context of holomorphic regularization the gauge kinetic function, or supersymmetric gauge coupling $f$, can be expressed as a Wilsonian coupling that depends on the conformal compensator \cite{kaplunovsky1994field,Giudice:1998xp,Arkani-Hamed:1998mzz}, as follows

\begin{equation}
    \label{eq:fgaugeholo}
    f(\varphi)=\frac{1}{g^2}+\,\frac{1}{8\pi^2}\Big(\delta\\ln{\frac{\Lambda\varphi}{\mu}}\,+\,\sum_fT(R_f)\ln{\mathcal{Z}_0}|_{\theta=0}\Big)=\frac{1}{g^2}+\frac{1}{8\pi^2}\Big[\delta-\sum_fT(R_f)\gamma_f\Big]F_{\varphi}\,\Big(\ln{\frac{\Lambda}{\mu}}\Big),
\end{equation}
where the quantity $\delta$ was specified in Section III, and on the right-hand side the first-order expansion of $\varphi=1+F_{\varphi}\theta^2$ was used.

On the other hand, one can perform an expansion up to order $\theta^2$ of the gauge kinetic function $f$, where the first–order component is the one associated with the gaugino mass parameter $m_{\lambda}$. This is because the function $f$ multiplies the field–strength term in the action (\ref{eq:LsuperWeyl}), giving rise to a mass term associated with the breaking of conformal symmetry

\begin{equation}
    \label{eq:fgaugemass}
    f=\frac{1}{g^2}\,+\,\theta^2\frac{m_{\lambda}}{g^2}\,+\,Im(f),
\end{equation}
comparing both expressions at order $\theta^2$, one concludes that

\begin{align}
\notag
    \frac{m_{\lambda}}{g^2}&\sim\frac{1}{8\pi^2}\Big[\delta-\sum_fT(R_f)\gamma_f\Big]F_{\varphi}\Rightarrow\\
    m_{\lambda}&\sim \frac{g^2}{8\pi^2}\Big[\delta-\sum_fT(R_f)\gamma_f\Big]F_{\varphi}\propto\frac{\beta_g}{g}\,F_{\varphi},
\end{align}

\begin{equation}
    m_{\lambda}\propto\frac{\beta_g}{g}\,F_{\varphi},
\end{equation}
which is the same result obtained in Eq. (\ref{eq:masg}) through the reduction of couplings, again with $F_{\varphi}\rightarrow M_0$.  This last relation, together with those in Eq. (\ref{eq:h&m}), makes it clear that, beyond inducing the conformal anomaly through the existence of a solution to the RoC condition in the soft-breaking sector, one can also derive explicit expressions for the soft parameters by analyzing the wave-function renormalization in superspace. These results coincide with those obtained previously through renormalization group invariance requirements and the reduction of couplings in Finite Unified Theories, thereby establishing a clear connection between both formalisms.

\section{Conclusions}

Throughout our analysis it has become clear that Zimmermann’s method of reduction of couplings is a very powerful tool when we are working with a model that possesses a large number of free parameters, since it yields relations among parameters that are invariant under the evolution with the energy scale. This is particularly useful when attempting to address some of the problems associated with the Standard Model, because with the help of the RoC one can construct physical theories (BSM models) that are free of ultraviolet divergences and at the same time reduce the number of free parameters by relating different sectors of the model (e.g. gauge and Yukawa sectors).

The theorem characterizing finite supersymmetric theories is one of the most powerful consequences of the method of  reduction of couplings, since it guarantees that the finiteness conditions preserve their form not only to all orders in perturbation theory, provided that a solution to equation (\ref{eq:solutionRGI}) exists, but also that they are independent of the renormalization scheme, ensuring that the relations among parameters are stable.

As we saw in Section IV, the idea of all-order finiteness provided by the RoC is in turn associated with an even larger symmetry in the dimensionless sector of the theory. As we showed, if in a supersymmetric Grand Unified model the conditions of the theorem are satisfied around a fixed point, then there will exist a conformal manifold associated with the dimensionless sector of the parameter space, such that the theory is superconformally invariant.
This implication of the finiteness theorem is not surprising, since at a more basic level there is already a one-loop scale invariance arising from the cancellation of the gauge and Yukawa beta functions, $\beta^{(1)}_g$ and $\beta^{(1)}_{C_{ijk}}$, respectively. However, this result generalizes to an all-order cancellation and culminates in the existence of a highly symmetric sector induced by the marginality of the superpotential operators that are consistent with the finite solution of the RoC.
This observation may also connect us, either directly or indirectly, to more fundamental theories that exhibit conformal regimes, such as those associated with certain classes of string theories \cite{Brignole:1995fb,Ibanez:1998xn} or even some versions of supergravity models \cite{Mavromatos:2012yu}, with finiteness acting as a bridge between these frameworks and physics at the scale of the Standard Model.

Likewise, in the search for a phenomenologically viable finite theory, Zimmermann’s method can be applied to the dimensionful soft-breaking sector in order to reduce the more than 100 additional free parameters introduced by the soft supersymmetry-breaking Lagrangian $\mathcal{L}_{\text{soft}}$. Thus, as has been well described in previous works \cite{Hisano:1997ua,Jack:1999aj}, after constructing Renormalization Group Invariants one obtains perturbatively stable relations for the soft parameters $m_i^2$, $h_{ijk}$, and $M_i$ in terms of the dimensionless supersymmetric couplings $g$ and $C_{ijk}$.
However, these relations exhibit an almost exact similarity to the form of the soft parameters in the anomaly-mediated supersymmetry-breaking (AMSB) scheme, except for the scalar-mass sum rule in our case, Eq. (\ref{eq:sumrule}). Thus, at some point in the derivation of these expressions within the context of the RoC, something happened to the possible tree-level contributions to SUSY breaking, such that only the one-loop contributions associated with the anomaly appear to dominate.
What we discovered in our study is that, in a supergravity theory whose action possesses conformal (Weyl) symmetry, the effect of reduction of couplings on a breaking scenario mediated by the gravity field $\varphi$ causes not only the conformal anomaly to manifest itself, but—given the structure of the action $\ref{eq:Lsugra}$—also leads to the cancellation of supersymmetry-breaking contributions arising from gravitational effects (GMSB) at tree level. This is the main reason why an AMSB-like pattern of SUSY breaking emerges in our finite Grand Unified models.

Another important aspect to highlight is that the breaking of supersymmetry through one-loop conformal anomaly contributions in a finite model evidently also breaks the conformal symmetry that was present at high energies in the dimensionless sector. However, thanks to the RGI invariant relations satisfied by the soft couplings at the breaking scale—particularly the scalar sum rule that cures the tachyonic mass spectrum problem \cite{Heinemeyer:2019vbc,Kobayashi:1997qx}—scale invariance remains manifest in this energy regime, resulting in a highly predictive model.

One consequence of this part of the analysis is the explicit form that the conformal Kähler potential must take for our all-loop finite model, since we found its structure coincides with the type of potentials studied in no-scale supergravity models. 
This observation is important for our attempt to relate our theories at the Grand Unification scale $M_{\text{GUT}}$ to more fundamental models. Moreover, the particular form of (\ref{eq:K22}) in the no-scale context \cite{Ellis:1983ew,Ellis:1984bm} is consistent with the constraint on gauge singlets given by the first finiteness condition (\ref{eq:C100}). In the same way, it exhibits a strong resemblance to the Kähler potential appearing in superconformal models used to study inflationary scenarios with $\alpha$-attractors \cite{Kallosh:2013yoa}, Starobinsky-like models \cite{Ellis:2015xna} and in inflation derived from $4D$ strings in more recent works  \cite{Antoniadis:2024ypf}. These two observations, pointing to a connection with supergravity (SUGRA) models at high energies and other type of more fundamental theories, are in full agreement with the hypothesis that in a divergence-free supersymmetric framework, the natural connection to a more fundamental theory arises through a no-scale scenario. 

With a view toward future work, it is expected that the cosmological consequences of a finite model with conformal symmetry, such as those discussed in this work, can be studied, since there are different approaches to analyzing them. One example is the work by Elizalde and  Odintsov in various studies \cite{Elizalde:1993qh,Elizalde:2015nya}, where a Starobinsky-type cosmological evolution pattern $f(R)$ \cite{Starobinsky:1980te,Antoniadis:2025pfa} is constructed starting from a finite model in curved spacetime. In this way, it should be possible to study the combined effect of curved-space contributions, as well as the effect of the symmetry under the non-compact group $SU(n,1)$ of the scalar kinetic terms, in the description of an inflationary model for some of the previously studied finite models that exhibit good phenomenology \cite{Heinemeyer:2019vbc}. Recently, an interesting example of a six dimensional $N=(1,0)$ higher-derivative gauge theory was constructed, which is one-loop finite \cite{Buchbinder:2026aub}. It would be interesting to find out if it is also possible to use the RoC method in this kind of higher dimensional setting.
 
\section*{Acknowledgements}

M.M. acknowledges useful discussions with G. Zoupanos.  Partially supported by UNAM project PAPIIT  IN111224 and SECIHTI project CBF2023-2024-548. LERR acknowledges SECIHTI for a graduate scholarship.

\bibliography{lms2eau.bbl}

\end{document}